\documentclass[english,aps,preprint]{revtex4}
\usepackage[T1]{fontenc}
\usepackage[latin9]{inputenc}
\usepackage{color}
\usepackage{babel}
\usepackage{amssymb}
\usepackage{graphicx}
\usepackage[unicode=true,pdfusetitle,
 bookmarks=true,bookmarksnumbered=false,bookmarksopen=false,
 breaklinks=false,pdfborder={0 0 1},backref=false,colorlinks=true]
 {hyperref}
\usepackage{breakurl}

\makeatletter
\@ifundefined{textcolor}{}
{%
 \definecolor{BLACK}{gray}{0}
 \definecolor{WHITE}{gray}{1}
 \definecolor{RED}{rgb}{1,0,0}
 \definecolor{GREEN}{rgb}{0,1,0}
 \definecolor{BLUE}{rgb}{0,0,1}
 \definecolor{CYAN}{cmyk}{1,0,0,0}
 \definecolor{MAGENTA}{cmyk}{0,1,0,0}
 \definecolor{YELLOW}{cmyk}{0,0,1,0}
}

\makeatother

\begin{document}

\preprint{This line only printed with preprint option}

\title{Growth and angular dependent resistivity of Nb$_{2}$Pd$_{0.73}$S$_{5.7}$
superconducting state single crystals fiber}

\author{Anil K. Yadav$^{\mbox{a,b}}$}

\email{anilsaciitb@gmail.com}

\author{Himanshu Sharma$^{\mbox{a}}$ }

\author{C. V. Tomy$^{\mbox{a}}$}

\email{tomy@phy.iitb.ac.in}

\author{Ajay D. Thakur$^{\mbox{c}}$}

\address{$^{\mbox{a}}$ Department of physics, Indian Institute of Technology
Bombay, Mumbai-400076 India}

\affiliation{$^{\mbox{b}}$ Department of Physics, Ch. Charan Singh University
Meerut, Meerut-250004 India }

\affiliation{$^{\mbox{c}}$ Department of Physics, Indian Institute of Technology
Patna, Patna-800013 India}
\begin{abstract}
We report the growth of Nb$_{2}$Pd$_{0.73}$S$_{5.7}$ superconducting
single crystal fibers via slow cooling solid state reaction method.
Superconducting transition temperature ($T_{c}\sim6.5$\,K) is confirmed
from magnetization and transport measurements. A comparative study
is performed for determination of superconducting anisotropy, $\Gamma$,
via conventional method (by taking ration of two superconducting parameters)
and scaling approach method. Scaling approach, defined within the
framework of the Ginzburg-Landau theory is applied to the angular
dependent resistivity measurements to estimate the anisotropy. The
value of $\Gamma$ close to $T_{c}$ from scaling approach is found
to be $\sim2.5$ that is slight higher compare to conventional approach
($\sim2.2$). Further, variation of anisotropy with temperature suggests
that it is a type of multi-band superconductor.
\end{abstract}
\maketitle

\section{introduction}

Ternary chalcogenide of non-superconducting compound Nb$_{2}$Pd$_{0.81}$Se$_{5}$
turns into a superconductor with superconducting transition temperature
$T_{c}\sim6.5$\,K when Se is replaced with S {[}1{]}. This superconductor
has caused a lot of interest to research community due to its extremely
large upper critical fields amongst the known Nb based superconductors
and shown a possibility to grow long flexible superconducting fibers
{[}1,2{]}. Structurally, this compound crystallizes in the monoclinic
structure with symmetry $C2/m$ space group {[}1,2,3{]}. Its structure
comprises laminar sheets, stacked along the b-axis, consisting of
Pb, Nb and S atoms. Each sheet contains two unique building blocks
of NbS$_{6}$ and NbS$_{7}$ atoms inter-linked by the Pd-atoms {[}1,3,4{]}.
Yu \emph{et al.}, have constructed the superconducting phase diagram
of Nb$_{2}$Pd$_{1-x}$S$_{5\pm\delta}$ ($0.6<x<1$ ) single crystal
fibers by varying composition of Pd and S and found maximum $T_{c}\sim7.43$\,K
in Nb$_{2}$Pd$_{1.1}$S$_{6}$ stoichiometry compound {[}2{]}. One
of the important parameter which needs to be determine precisely for
this compound is the anisotropy ($\Gamma$) as it shown extremely
large direction dependent upper critical field {[}1{]}. In the conventional
approach, the anisotropy is determined as ration of two superconducting
parameters (such as band dependent effective masses, penetration depth,
upper critical fields etc.) in two orientations w.r.t. the crystallographic
axes and applied magnetic field {[}5{]}. Zhang \emph{et al.}, {[}1{]}
have determined the temperature dependent anisotropy in this compound
using the above conventional method by taking the ratio of $H_{c2}(T)$
in two orientations. However, in this case, estimation of $H_{c2}(0)$
is subject to different criteria and formalism which may introduce
some uncertainty in the anisotropy ($\Gamma$) calculation {[}6{]}.
Blatter \emph{et al.}, have given a simple alternate way to estimate
the anisotropy of a superconductor, known as the scaling approach
{[}7{]}. In this approach, any anisotropic data can be changed into
isotropic form by using some scaling rule in which only one parameter
has to adjust for which all isotropic curves collapse into single
curve, that adjusted parameter is anisotropy of superconductor. Thus
its limits the uncertainty in the determination of $\Gamma$ as compared
to the conventional approach. Employing scaling approach, Wen et al.,
have estimated the anisotropy of several Fe-based superconductors
such as NdFeAsO$_{0.82}$F$_{0.18}$ {[}8{]}, Ba$_{1-x}$K$_{x}$Fe$_{2}$As$_{2}$
{[}6{]} and Rb$_{0.8}$Fe$_{2}$Se$_{2}$ {[}9{]}. Shahbazi \emph{et
al.}, have also performed similar studies on Fe$_{1.04}$Se$_{0.6}$Te$_{0.4}$
{[}10{]} and BaFe$_{1.9}$Co$_{0.1}$As$_{2}$ {[}11{]} single crystals.
In this paper, we report the anisotropy estimation of Nb$_{2}$Pd$_{0.73}$S$_{5.7}$
single crystals via conventional and scaling approach methods near
$T_{c}$. We also provide further evidence that the bulk superconducting
anisotropy is not universally constant, but is temperature dependent
down to $T_{c}$.

\section{method}

Single crystal fibers of Nb$_{2}$Pd$_{0.73}$S$_{5.7}$ were synthesized
via slow cooling of the charge in the solid state reaction method,
as reported in reference {[}1{]}. Starting raw materials (powder)
Nb (99.99\%), Pd (99.99\%) and S (99.999\%) were taken in the stoichiometry
ratio of 2:1:6 and mixed in an Ar atmosphere inside a glove box. The
well-homogenized powder was sealed in a long evacuated quartz tube
and heated to 800ïC with a rate of 10$^{\circ}$C/h. After the reaction
for 24 hours at this temperature, the reactants were cooled down at
a rate of 2$^{\circ}$C/h to 360$^{\circ}$C, followed by cooling
to room temperature by switching the furnace off. As-grown samples
look like a mesh of small wires when viewed under an optical microscope.
Some part of the as-grown sample was dipped in dilute HNO$_{3}$ to
remove the bulk material and to pick up a few fiber rods for further
measurements. X-ray diffraction (XRD) was performed on powdered Nb$_{2}$Pd$_{0.73}$S$_{5.7}$
single crystal fibers for structure determination. High energy x-ray
diffraction analysis (EDAX) is used to identify the chemical elements
and composition. Magnetization measurement was performed using a superconducting
quantum interference device - vibrating sample magnetometer (SQUID-VSM,
Quantum Design Inc. USA). Angular dependent resistivity was carried
out using the resistivity option with horizontal rotator in a physical
property measurement system (PPMS) of Quantum Design Inc. USA. Electrical
connections were made in four probe configuration using gold wires
bonded to the sample with silver epoxy.

\section{Results}

\subsection{Structure analysis}

\begin{figure}[h]
\begin{centering}
\includegraphics[scale=0.55]{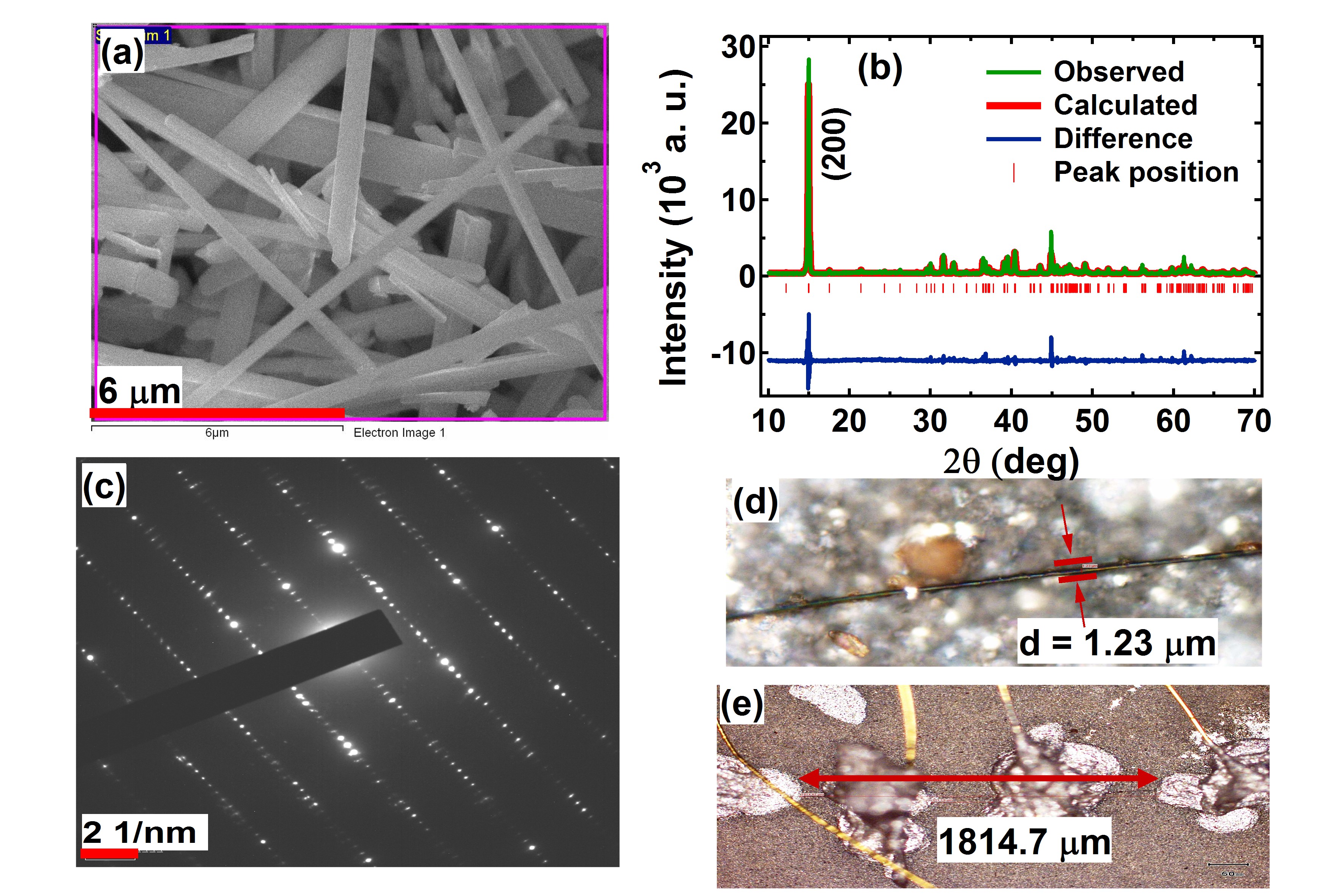}
\par\end{centering}

\caption{ \label{fig1} (Color online) (a) SEM image of bunch of single crystal
fibers of Nb$_{2}$Pd$_{0.73}$S$_{5.7}$. (b) X-ray diffraction patterns:
observed (green), calculated (red) and difference (blue) (c) SAED
pattern of a single crystal fiber (d) optical image of a typical cylindrical
wire used for transport study (e) Four probe connections on a fiber.}
\end{figure}

Figure \ref{fig1}(a) shows the scanning electron microscope (SEM)
image of Nb$_{2}$Pd$_{0.73}$S$_{5.7}$ single crystals fibers. It
is clear from the image that the fibers are grown in different shapes
and lengths. Figure \ref{fig1}(b) shows the XRD patterns of powdered
Nb$_{2}$Pd$_{0.73}$S$_{5.7}$ single crystals. Rietveld refinement
was performed on the powder XRD data using $C2/m$ monoclinic crystal
structure of Nb$_{2}$Pd$_{0.81}$Se$_{5}$ as reference in the FullProf
suite software. The lattice parameters ($a$= 12.154(1)A, $b$ = 3.283(7)A
and $c$ = 15.09(9)A) obtained from the refinement are approximately
same as reported earlier in reference {[}1,3{]}, even though the intensities
could not be matched perfectly. Peak (200) is found to be the one
with the highest intensity even when the XRD was obtained with a bunch
of fibers, indicating a preferred crystal plan orientation along the
($l$00) direction in our powdered samples. Similar preferred orientation
was also reported for single crystals in reference {[}2{]}. This may
be the reason for the discrepancy in the intensities between the observed
and the fitted XRD peaks. Further, to confirm the single crystalline
nature of the fibers, we have taken the selective area electron diffraction
(SAED) pattern of the fibers; a typical pattern is shown in Figure
\ref{fig1}(c). Nicely ordered spotted diffraction pattern confirms
the single crystalline nature of the fibers. Figure \ref{fig1}(d)
shows the optical image of a typical cylindrical fiber of diameter
$\sim1.2\,\mbox{\ensuremath{\mu}m}$ and of length $\sim1814\,\mbox{\ensuremath{\mu}m},$
which was used for the four probe electrical resistivity measurements
(Fig. \ref{fig1}(e) shows the gold wires and silver paste used for
the electrical connections). All chemical elements are found to be
present in the compound with slight variation from starting composition
in EDAX analysis.

\subsection{Confirmation of superconducting properties}

\begin{figure}[h]
\begin{centering}
\includegraphics[scale=0.55]{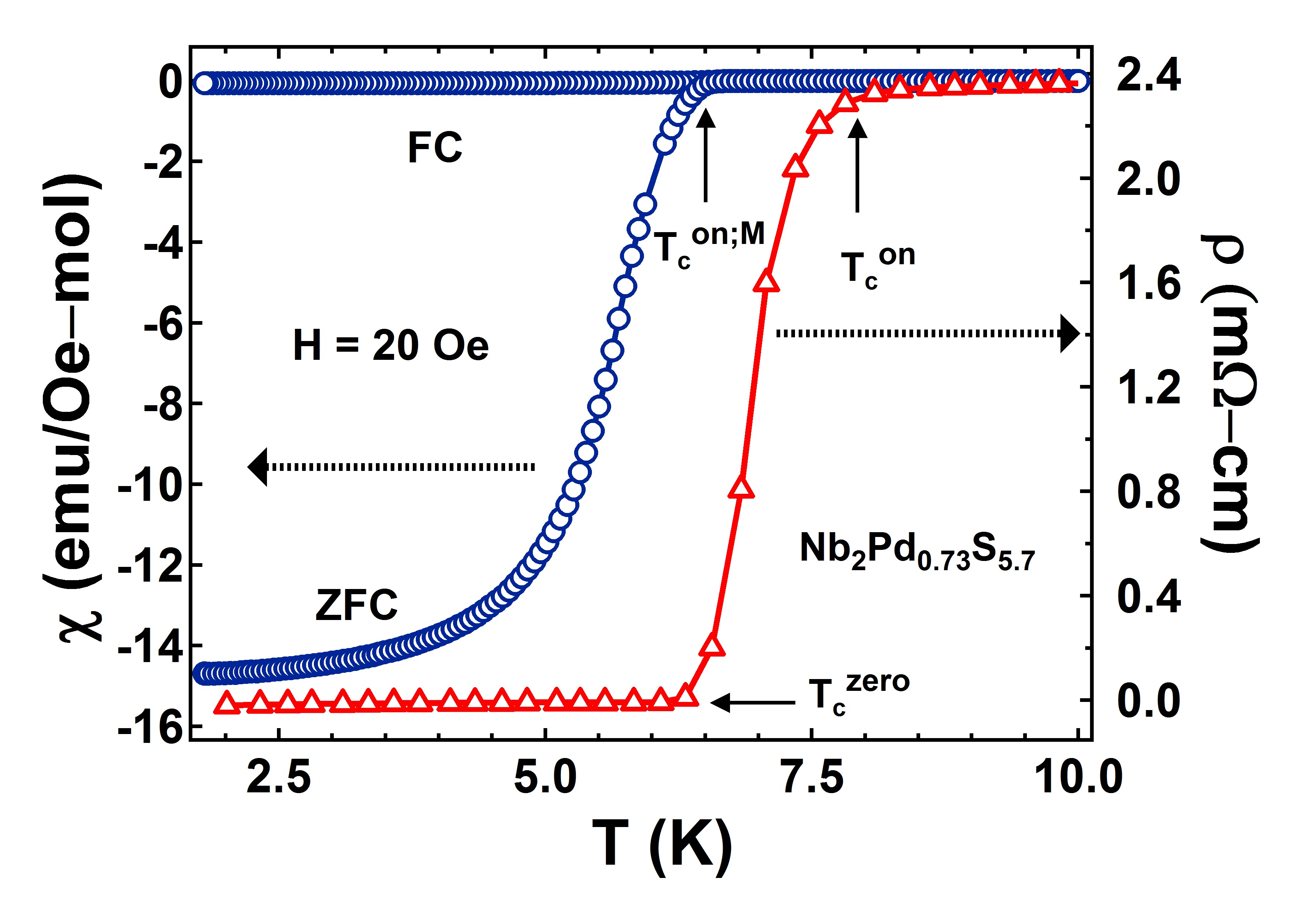}
\par\end{centering}

\caption{\label{fig2}(Color online) Zero field-cooled (ZFC) and field-cooled
(FC) magnetization curves at 20\,Oe (open circle) and resistivity
measurement at zero field (open triangle). Onset superconducting transition
temperature, $T_{c}^{{\rm on;M}}$, from magnetization and zero resistivity
transition temperature, $T_{c}^{{\rm zero}}$, from resistivity measurements
confirm the $T_{c}$ of Nb$_{2}$Pd$_{0.73}$S$_{5.7}$ superconductor.}
\end{figure}

In order to confirm the occurrence of superconductivity in the prepared
single crystals, magnetic measurement was performed on a bunch of
fibers (as alone single crystal fiber did not give large enough signal
in magnetization). Figure \ref{fig2} shows a part of the temperature
dependent zero field-cooled (ZFC) and field-cooled (FC) magnetization
measurements at H = 20Oe. The onset superconducting transition temperature
($T_{c}^{{\rm on;M}})$ is observed to be $\sim6.5$\,K which is
taken from the bifurcation point of ZFC and FC curves. In order to
confirm the superconducting nature of the grown single crystal fibers,
resistivity was measured using one of the fibers removed from the
ingot. We have plotted a part of resistivity measurement (in zero
applied magnetic fields) in Fig. \ref{fig2} along with the magnetization
curve where zero resistivity transition temperature, $T_{c}^{{\rm zero}}$
matches well with the onset transition temperature of magnetization,
$T_{c}^{{\rm on;M}}$ as well as the $T_{c}$ reported in reference
{[}1,2{]}. However, the onset transition temperature from resistivity
($T_{c}^{{\rm on}}$ : the temperature at which resistivity drop to
90\% from normal state resistivity) is found to be $\sim7.8$\,K,
which is comparable to the optimized maximum $T_{c}^{{\rm on}}$ for
this compound reported by Yu \emph{et al.}, {[}2{]}. The narrow superconducting
transition width ($\sim1.3$\,K) in resistivity indicates the quality
of the single crystal fibers (see Fig. \ref{fig2}). The residual
resistivity ratio $(RRR\thickapprox\frac{R(300\,{\rm K)}}{R(8\,{\rm K)}})$,
which indicates the metallicity of a material, is found to be $\sim3.4$
for our sample. This value of RRR is much less than the corresponding
value for good conductors, that categorized it as bad metals.

\subsection{Angular dependent transport properties}

\begin{figure}[h]
\begin{centering}
\includegraphics[scale=0.45]{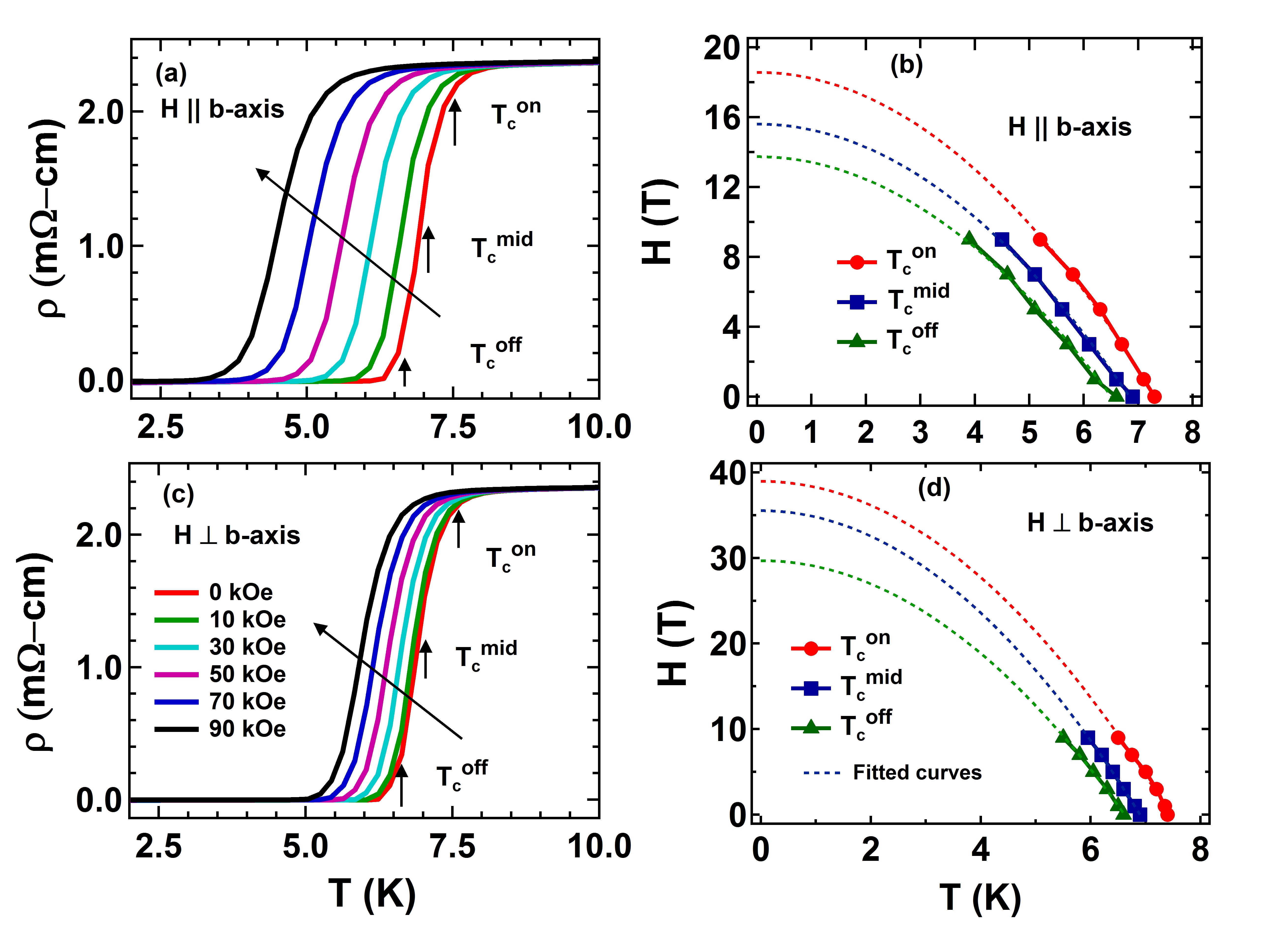}
\par\end{centering}

\caption{\label{fig3}(Color online) Temperature dependent resistivity plots
at different applied fields vary from 0\,kOe to 90\,kOe (a) for
H || b-axis (c) for H $\bot$ b-axis. (b) and (d) plots show H--T
phase diagrams at $T_{c}^{{\rm on}}$, $T_{c}^{{\rm mid}}$ and $T_{c}^{{\rm off}}$
transition temperatures. Dashed curves show the fitting curves corresponding
empirical formula, $H_{c2}(0)=H_{c2}(T)(1-(T/T_{c})^{2})$. }
\end{figure}

In order to estimate the superconducting anisotropy properties, we
have to assign single crystal fibers orientation axis. Since we cannot
assign a growth direction for our cylindrical single crystal fibers
from XRD due to very fine single crystals therefore we have adopted
b-axis along length of fibers as given in reference {[}1{]} because
of same synthesis method is followed to grow these single crystal
fibers. Figures \ref{fig3}(a) and \ref{fig3}(c) show the resistivity
plots as function of temperature in different applied magnetic fields
from zero to 90\,kOe along H || b-axis and H $\bot$ b-axis. Three
transition temperatures, $T_{c}^{{\rm on}}$, $T_{c}^{{\rm mid}}$and
$T_{c}^{{\rm off}}$ are marked in the figure using the criteria,
90\%$\rho_{n}$, 50\%$\rho_{n}$ and 10\%$\rho_{n}$ (where $\rho_{n}$
is normal state resistivity at 8\,K), respectively. The $T_{c}$
shifts toward the lower temperatures as field increases with the rate
of 0.05\,K/kOe and 0.02\,K/kOe along H || b-axis and H $\bot$ b-axis,
respectively. The H\textendash{}T phase diagrams are plotted at three
transition temperatures in Figs. \ref{fig3}(b) and \ref{fig3}(d)
for both the orientations. In order to find out upper critical fields
($H_{c2}(0)$), these H\textendash{}T curves are fitted with the empirical
formula, $H_{c2}(0)=H_{c2}(T)(1-(T/T_{c})^{2})$ {[}1, 2{]}, further
these fitted curves have been extrapolated to the zero temperature
to extract the $H_{c2}(0)$ values, that come out to be $\sim$180\,kOe
and $\sim$390\,kOe at $T_{c}^{{\rm on}}$ along H || b-axis and
H $\bot$ b-axis, respectively. Conventionally the anisotropy is found
to be $\sim2.2$, estimated by taking ratio of $H_{c2}(0)$ values
in two orientations. In order to corroborate the $\Gamma$ values
further, we have measured the angular dependent resistivity $\rho(\theta)$
at different magnetic fields at certain temperatures close to $T_{c}$.

\begin{figure}[h]
\begin{centering}
\includegraphics[scale=0.45]{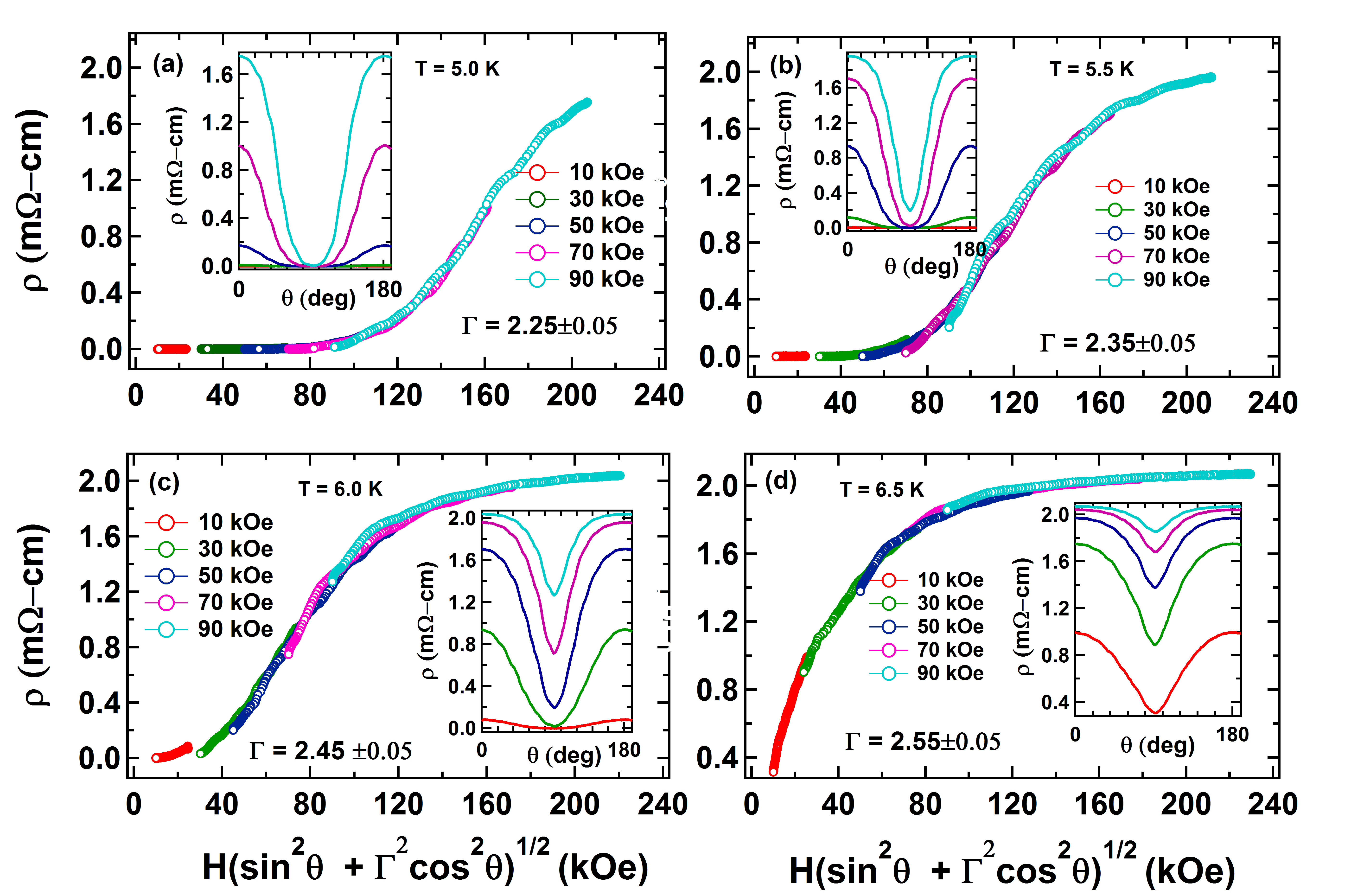}
\par\end{centering}

\caption{\label{fig4}(Color online) Insets of figure (a) to (d) show resistivity
($\rho$) plots as a function of angle, $\theta$ (angle between b-axis
and applied magnetic fields) at fields 10\,kOe, 30\,kOe, 50\,kOe,
70\,kOe and 90\,kOe for temperatures (a) 5\,K (b) 5.5\,K (c) 6.0\,K
and (d) 6.5\,K and main panels of figure show the resistivity plots
as function of scaling field $\tilde{H}$ = $H\,\sqrt{{\rm sin^{2}}\theta+\Gamma^{2}{\rm cos^{2}}\theta}$.}
\end{figure}

\begin{figure}[h]
\begin{centering}
\includegraphics[scale=0.45]{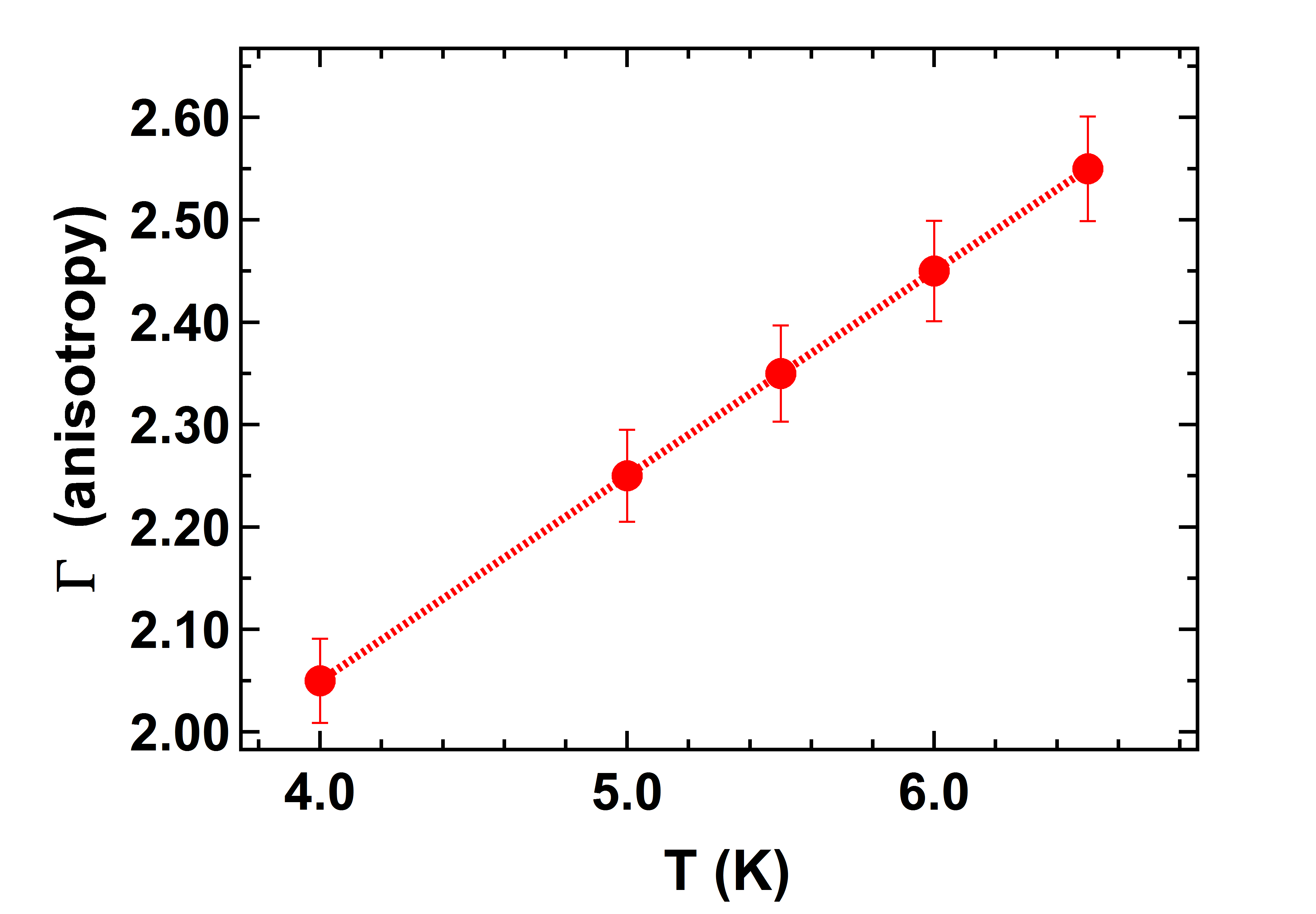}
\par\end{centering}

\caption{\label{fig5}(Color online) Anisotropy variation with temperature
measured from angular dependent resistivity.}
\end{figure}

The insets of Figs. \ref{fig4}(a), (b), (c) and (d) show $\rho(\theta)$
curves at 10\,kOe, 30\,kOe, 50\,kOe, 70\,kOe and 90\,kOe for
T = 5.0\,K, 5.5\,K, 6.0\,K and 6.5\,K, respectively. All the $\rho(\theta)$
curves show a symmetric dip at $\theta=90$$^{\circ}$ and a maximum
at 0$^{\circ}$ and 180$^{\circ}$. In all the curves, the center
of the dip shifts from zero to non-zero resistivity as the temperature
and field increases. The main panel of the Fig. \ref{fig4} shows
rescaled $\rho(\theta)$ curves of 10\,kOe, 30\,kOe, 50\,kOe, 70\,kOe
and 90\,kOe fields at temperatures (a) 5.0\,K (b) 5.5\,K (c) 6.0\,K
and (d) 6.5\,K, respectively using the rescaling function: 
\begin{equation}
\tilde{H}=H\,\sqrt{{\rm sin^{2}}\theta+\Gamma^{2}{\rm cos^{2}}\theta}
\end{equation}
 where $\Gamma$ is anisotropy and $\theta$ is angle between the
field and crystal axis. All rescaled curves at fixed temperature are
now isotropic, i.e., all curves collapse on the single curve. In this
method only anisotropic parameter, $\Gamma$, was adjusted to convert
data into the isotropic form, that value of $\Gamma$ is the anisotropy
at that temperature. 
 Figure \ref{fig5} shows temperature dependent anisotropy ($\Gamma(T)$)
plot which is obtained from the angular resistivity data. Anisotropy
decreases slowly as the temperature goes down in superconducting state.
As Zhang \emph{et al.} {[}1{]} have explained that this dependency
of anisotropy in temperature may be due to the opening of superconducting
gap of different magnitude on different Fermi surface sheets where
each associated with bands of distinct electronic anisotropy. Li \emph{et
al.}, have reported similar temperature dependent anisotropy behavior
for Rb$_{0.76}$Fe$_{2}$Se$_{1.6}$ , Rb$_{0.8}$Fe$_{1.6}$Se$_{2}$
, Ba$_{0.6}$K$_{0.4}$Fe$_{2}$As$_{2}$, Ba(Fe$_{0.92}$Co$_{0.08})_{2}$As$_{2}$
single crystals and explained that this may be due to the multiband
effect or gradual setting of pair breaking due to spin-paramagnetic
effect {[}9{]}. Shahbazi \emph{et al.}, have also reported similar
results for Fe$_{1.04}$Te$_{0.6}$Se$_{0.4}$ and BaFe$_{1.9}$Co$_{0.8}$As$_{2}$
single crystal through angular dependent transport measurements {[}10,11{]}.
Various theoretical models for study of Fermi surface have supported
the presence of multiband superconducting gap in Fe-based superconductors
{[}12,13,14{]}. Here, the density functional theory (DFT) calculation
indeed has shown that the Nb$_{2}$Pd$_{0.81}$S$_{5}$ superconductor
is a multi-band superconductor {[}1{]}. Compared to MgB$_{2}$ {[}15,16{]}
and cuprate superconductors {[}17{]} the anisotropy of Nb$_{2}$Pd$_{0.73}$S$_{5.7}$
is very small; however, it is comparable with some of the iron based
(Fe-122 type) superconductors {[}9{]}.

\section{Conclusions}

In conclusion, we have successfully synthesized the Nb$_{2}$Pd$_{0.73}$S$_{5.7}$
single crystal fibers via slow cooling solid state reaction method.
Superconducting properties of sample have been confirmed via magnetic
and transport measurements. Conventionally, upper critical fields
are measured from magneto-transport study. Angular dependence of resistivity
are measured in presence of magnetic fields at different temperatures
in superconducting state which further rescaled using a scaling function
to convert isotropic form that direct provides anisotropy. The anisotropy
is found to be $\sim2.5$ near $T_{c}$ which is less $\sim2.2$ compare
to achieve from conventional method. Anisotropy decreases slowly with
decreasing temperature, which is attributed to the multi-band nature
of the superconductor.
\begin{acknowledgments}
AKY would like to thank CSIR, India for SRF grant.\end{acknowledgments}

\end{document}